\documentclass[
    ,final            
  ]
  {aipproc}

\layoutstyle{6x9}

\DeclareGraphicsExtensions{.eps,.ps} 

\newcommand{\LHC}{LHC}

\newcommand{\TEVATRON}{Tevatron}

\newcommand{\ATLAS}{ATLAS}
\newcommand{\PYTHIA}{PYTHIA}
\newcommand{\pdf}{PDF}

\newcommand{\NLOa}{NLO}

\newcommand{\lumi}{\mbox{${\cal L}$}}

\newcommand{\aTGC}{anomalous~TGC}
\newcommand{\TGC}{TGC}
\newcommand{\OpOb}{OO}  

\newcommand{\lgamma}{\lambda_\gamma}
\newcommand{\dkgamma}{\Delta\kappa_\gamma}

\newcommand{\dgOneZ}{\Delta g^1_Z}
\newcommand{\lz}{\lambda_Z}
\newcommand{\dkz}{\Delta\kappa_Z}

\newcommand{\LFF}{\Lambda_\txt{FF}}

\newcommand{\txt}[1]{{\mathrm{#1}}} 

\newcommand{\pT}[1]{P^T_\txt{#1}}        

\newcommand{\mass}[1]{M_\txt{#1}}
\newcommand{\pseudorapitidy}[1]{\eta_\txt{#1}} %
\newcommand{\mycos}{\cos \theta_{\bar{q},\gamma^\star}}

\newcommand{\deltaR}[2]{\Delta R(\mathrm{#1,#2})}

\newcommand{\sumPtJet}{ \sum_\txt{jets} \vec{\pT{jet}}_i }

\newcommand{\lzSystStatLimit}{
        -0.0065_\txt{stat.},~ & -0.0032_\txt{syst.} & 
                < \lz < 
        & +0.0066_\txt{stat.},~ & +0.0031_\txt{syst.} 
}
\newcommand{\lzCombinedLimit}{ -0.0073  <    \lz  < 0.0073 }

\newcommand{\dkzSystStatLimit}{
        -0.10_\txt{stat.},~ & -0.024_\txt{syst.} & 
                < \dkz < 
        & +0.12_\txt{stat.},~ & +0.024_\txt{syst.} 
}
\newcommand{\dkzCombinedLimit}{ -0.11   < \dkz < 0.12 }

\newcommand{\dgOneZSystStatLimit}{
        -0.0064_\txt{stat.},~ & -0.0058_\txt{syst.} & 
                < \dgOneZ < 
        & +0.010_\txt{stat.},~ & +0.0058_\txt{syst.} 
}
\newcommand{\dgOneZCombinedLimit}{ -0.0086 < \dgOneZ < 0.011 }
\newcommand{\lgammaSystStatLimit}{
        -0.0033_\txt{stat.},~ & -0.0012_\txt{syst.} & 
                < \lgamma <
        & +0.0033_\txt{stat.},~ & +0.0012_\txt{syst.} 
}
\newcommand{\lgammaCombinedLimit}{ -0.0035      < \lgamma < 0.0035 }

\newcommand{\dkgammaSystStatLimit}{
        -0.073_\txt{stat.},~ & -0.015_\txt{syst.} & 
                < \dkgamma < 
        & +0.076_\txt{stat.},~ & +0.0076_\txt{syst.} 
}
\newcommand{\dkgammaCombinedLimit}{ -0.075 < \dkgamma < 0.076 }

\begin{document}

\title{Prospects for Probing Triple Gauge-boson Couplings at the LHC}

\author{Matt Dobbs}{
  address={ \texttt{<Matt.Dobbs@cern.ch>},
    Physics~Division, Lawrence~Berkeley~National~Laboratory, 
           1~Cyclotron~Road, Berkeley, USA~94720}
}

%

\begin{abstract}

 In these proceedings I explore one aspect of gauge-boson physics at
 the LHC---Triple Gauge-boson Couplings
 (\TGC s) in $WZ$ and $W\gamma$ production. Methods for
 extracting confidence limits on anomalous \TGC s are assessed, while
 accounting for the effects of higher order QCD corrections and
 contributions from other theoretical and detector related
 systematics. Detector response has been parametrised according to the
 ATLAS detector's specifications.  
 A strategy for reporting the anomalous coupling limits is introduced
 which removes the ambiguities of form factors by reporting the
 limits as a function of a cutoff operating on the diboson system
 invariant mass. Techniques for measuring the energy dependence of
 anomalous couplings are demonstrated.

\end{abstract}

\maketitle


\section{Introduction}

 The HCP 2004 talk associated with these proceedings covered the more
 general topic of gauge-boson physics at the LHC. I reviewed prospects
 for the measurement of the W-mass~\cite{Atlas:TDR,marques:2004Wmass}, 
 the electroweak
 mixing angle from the forward-backward asymmetry in dilepton
 production~\cite{Sliwa:2000}, Triple Gauge-boson Couplings in diboson
 production, and the production of three gauge
 bosons~\cite{Bell:2003phd}.
I highlighted some of the many challenges associated with making such
 measurements, including the simulation of processes for which accurate Monte Carlo
 predictions are missing (see Ref.~\cite{Dobbs:2004qw} for a review of
 relevant event generator techniques) and the need for supporting
 measurements of parton density functions and luminosity.

 I have chosen to focus these proceedings on the topic of Triple
 Gauge-boson Couplings in $W\gamma$ and $WZ$ production, since most of
 the other topics appear in published form elsewhere, and recent \TGC\
 results have not yet appeared in the public domain. More detailed
 descriptions of these \TGC\ studies can be found in \ATLAS\ Internal
 Notes \cite{Dobbs:2002WZ,Dobbs:2002WA} and Ref.~\cite{Dobbs:2002phd}.
 These processes have been studied in the context of the CMS detector
 in
 Ref.~\cite{mackay:1998phd,Muller:2000tgc,Mackay:2001tgc1,Mackay:2001tgc2}.

\subsection{Triple Gauge-boson Couplings}   

In the Standard Model (SM), the gauge-bosons interact not only with
matter particles, but also with one another.  These interactions
manifest themselves as couplings between three (or more) gauge-bosons,
such as a $WWZ$ or $WW\gamma$ coupling, referred to as triple
gauge-boson couplings (\TGC's). The existence of these couplings has
been beautifully verified at
LEP~\cite{Barate:1999gu,Abreu:1999ra,Acciarri:1998aq,Abbiendi:1998bg}.
\TGC s are tightly connected with the symmetry properties of the SM
and reflect the full mathematical gauge group structure of the
fundamental interactions. This gauge structure produces cancellations
in the production of $W^+W^-$ and $WZ$ pairs. Without these
cancellations, the cross section for longitudinally polarised $W^+W^-$
and $WZ$ pairs would grow proportional to the diboson invariant
mass squared, violating unitarity at relatively low energies. Because
these cancellations are so important for the consistency of the model,
it is necessary to test them at the highest accuracy possible. The
production of gauge-boson pairs in hadronic collisions provides a
direct test of these couplings.  While the $pp\to W^+W^-$ mechanism
receives contributions from both the $WWZ$ and $WW\gamma$ coupling,
the $pp\to WZ$ and $pp\to W\gamma$ channels allow for the
direct independent measurement of the $WWZ$ and $WW\gamma$ couplings
respectively. Other gauge-boson self interactions such as $ZZZ$,
$ZZ\gamma$, $Z\gamma\gamma$, and $\gamma\gamma\gamma$ vertices are not
allowed in the Standard Model, because neither the $Z$ nor the
$\gamma$ carries charge or weak isospin which are the quantum numbers
to which the gauge-bosons couple (anomalous \TGC s in $pp\to ZZ$ and
$pp\to Z\gamma$ have been studied in the context of the \LHC\ in
Refs.~\cite{Muller:2000tgc,Mackay:2002tgc,Hassani:2002ZZ,Hassani:2002ZA}).
Vertices containing an odd number of $W$-bosons ($WZZ$,
$W\gamma\gamma$, $WZ\gamma$, $WWW$) are excluded by charge
conservation. The self interactions also encompass interactions
between four gauge-boson (quartic couplings).


The most general Lorentz and gauge invariant anomalous $WWZ$ vertex
which conserves charge and parity is described by an effective
Lagrangian with 3 {\it model independent} anomalous \TGC\ parameters
$\dgOneZ,~\dkz,~\mbox{and}~\lz$. The corresponding $WW\gamma$ vertex
is described by 2 parameters $\dkgamma,~\mbox{and}~\lgamma$.  These
parameters are all zero in the SM, and strictly speaking may be
energy dependent (see discussion below). Experimental attempts to measure anomalous
\TGC\ parameters probe the low energy remnants of new physics which
may be operating at a much higher energy scale. Measurements of this
type would be most interesting in the scenario where direct searches
for new particles which affect the gauge-boson interactions fail to
observe any substantial deviation from the SM.

The study described in this paper is optimised for ``low luminosity''
(10$^{33}$cm$^{-2}$s$^{-1}$) \LHC\ conditions. It focuses on the
$pp\to W\gamma \to l^\pm \nu \gamma$ and $pp\to WZ \to l^\pm
\nu l^+ l^-$ processes (where $l^\pm$ is an electron or muon).
Detector effects have been included in the form of a fast
parametrisation~\cite{RichterWas:1998atlfast} of the \ATLAS\ detector
response. Next-to-leading order (\NLOa) QCD corrections to diboson
production are large at \LHC\ energies, particularly in the physically
interesting region of high transverse momentum which is the
region of maximum sensitivity to \aTGC's. These effects have been
accounted for using the \NLOa\ Baur, Han, and Ohnemus (BHO)
generators~\cite{Baur:1993ir,Baur:1994aj}. The BHO generators have
been modified to provide event weights as a function of the
anomalous coupling parameters, as discussed in the appendix of
Ref.\cite{Dobbs:2002WA}. For events with a coloured parton in the
final state, \PYTHIA~6.136~\cite{Sjostrand:2001wi} is used for
independent fragmentation and subsequent hadronization of the coloured
parton.\footnote{ The standard parton shower approach cannot be
applied to the events produced by the BHO generator, because this
would double count regions of phase space. See
Ref.~\cite{Dobbs:2004qw} for a description of recent advances in this
subject.} \PYTHIA~6.136 has been used for the simulation of the
background processes, with a single constant $k$-factor of 1.5 applied
to roughly account for the effect \NLOa\ corrections might have on the
total background rate. Background rates and shapes have a relatively
small impact on the confidence limits reported in this paper. As such,
background simulations using next-to-leading order matrix elements are
not expected to change the results significantly.


\section{Backgrounds and Event Selection}

The $WW\gamma$ and $WWZ$ vertices will be probed at \LHC\ using the
muon and electron decay channels, $pp\rightarrow W\gamma
\rightarrow l^\pm \nu \gamma$ and $pp\rightarrow WZ \rightarrow
l^\pm \nu l^+ l^-$. These processes provide striking detector
signatures consisting of high transverse momentum charged leptons
and/or a photon, together with missing transverse momentum. Events can
be triggered either with the single muon, single electron, and/or the
high $P_T$ photon triggers.  Hadronic decay channels are difficult to
separate from QCD backgrounds, and the addition of these channels are
not expected to significantly improve the precision of the
measurements.

The kinematic selection criteria for this analysis have been optimised
not only to maximise the signal significance, but also to minimise the
effect of systematics and to maximise the sensitivity to
\aTGC s. For the purpose of optimising these cuts, a leading order
signal simulation with showering and hadronization is used. This
avoids the possibility that the event selection makes use of the
differences in the simulation methods that have been used for signals
and backgrounds (e.g.\ a cut on the number of jets would take advantage
of the fact that the \NLOa\ signal simulation can produce only one
final state hard jet).

Several backgrounds will mimic the $W\gamma$ signal. The most
important background processes are: \\
\underline{$W(\rightarrow\tau\nu)\gamma$ with leptonic tau decays}
  This process is also sensitive to the \TGC\ vertex, but is considered
  a background for the purposes of this study since the $\tau$'s are
  more difficult to reconstruct. The contribution from this process will
  be reduced by lepton transverse momentum cuts, because the secondary
  charged leptons from the $\tau$-decay will have reduced transverse
  momentum as compared to the direct lepton from the $W$-decay. At
  \TEVATRON\ energy, this effect renders the leptonic $\tau$ decay
  background negligible~\cite{kelly96}. This is not the case at \LHC\
  energy. \\
\underline{$Z^0 \gamma$ production with leptonic decays}, 
  with one charged lepton escaping detection. \\
\underline{Heavy flavours $t\bar{t}(\gamma)$ and $b\bar{b}(\gamma)$}
  Though the signature for these events is very different from the
  signal, the cross section is so high that the tails of these
  distributions become important. The primary defence against these
  events is a simple jet veto. It may be possible to further improve 
  the selection by rejecting those events in which a top quark can be 
  reconstructed. \\
\underline{$Z^0+$jet and $W+$jet production}, with the jet
  mis-identified as a photon. This is the most challenging
  background. Its contribution will depend strongly on the particle 
  ID capabilities of the detector.

The selection criteria for the $W\gamma$ analysis are: 
(1) one isolated photon with 
  $\pT{\gamma}>100~\mathrm{GeV},~|\pseudorapitidy{\gamma}|<2.5$ 
  and no other reconstructed photon with 
  $\pT{\gamma}>80~\mathrm{GeV},~|\pseudorapitidy{\gamma}|<2.5$, 
(2) one isolated electron or muon,
  $\pT{l^\pm}>25~\mathrm{GeV},~|\pseudorapitidy{l^\pm}|<2.5$ 
  and no other charged lepton with
  $\pT{l^\pm}>20~\mathrm{GeV},~|\pseudorapitidy{l^\pm}|<2.5$,
(3) missing transverse momentum $\pT{miss}>25~\mathrm{GeV}$,
(4) vector sum of jet transverse momenta (jet veto) 
    $\sumPtJet<100~\mathrm{GeV}$,
(5) charged lepton to photon separation
  $\Delta R(l^\pm,\gamma)~=~\sqrt{ \Delta\phi^2 + \Delta\eta^2 }>1$,
and 
(6) a solution for the neutrino longitudinal momentum exists which is 
   consistent with it arising from a $W$.
The effect of these cuts on the signal and background rates are
tabulated in Table~\ref{t_selection_WA}. About 6900 event candidates 
will be observed with an integrated luminosity of 30~fb$^{-1}$, 2600 
of which will be background.
\begin{table}
\begin{small}
\begin{tabular}{|l|ccccc|c|c|} \hline
		& 
		& 
		& 
		& 
		& $W\gamma\rightarrow$
                & all
                & $W\gamma$
\\
		& $Z\gamma$
		& $W+$jet
		& $Z+$jet
		& $t\bar{t}(\gamma)$ 
		& $\tau\nu\gamma$ 
                & Backgrd
                & Signal
\\ \hline
$\pT{\gamma}>100$ GeV
		& 1277
		& 2097
		& 2101
		& 945  
		& 665
		& 8153
		& 10638
\\
$\pT{l^\pm}>25$ GeV 
		& 1196
 		& 1938
 		& 1800
		& 837 
 		& 586
 		& 7098
		& 10066
\\
$\pT{miss}>25$ GeV         
 		& 377
		& 1557
 		& 215
 		& 689 
 		& 574
 		& 3511
		& 7311
\\
$\deltaR{\gamma}{l^\pm}>$1
 		& 376
		& 1543
 		& 183
 		& 611 
 		& 574
 		& 3385
		& 6791
\\
$\sumPtJet<100$ GeV        	
 		& 341
		& 1280
 		& 133
 		& 286 
 		& 534
 		& 2623
		& 4262
\\ \hline
\end{tabular} 
\end{small}

\caption
{\label{t_selection_WA} The number of events surviving after each of
the kinematic cuts for the $W\gamma$ analysis is applied cumulatively
for an integrated luminosity $\lumi=30~\mathrm{fb}^{-1}$.  The
$\gamma$+jet, $b\bar{b}(\gamma)$, and $W\rightarrow l \nu \gamma$
processes have been included in the background totals, but are not
shown individually in the table.  }
\end{table}


\begin{table}
\begin{small}
\begin{tabular}{|l|ccc|c|c|} \hline
                & $Z+$jet       
		& $ZZ$
		& $t\bar{t}$   
                & All Backgrd
                & $WZ$ Signal
\\ \hline
3 leptons, $\pT{l^\pm}>25$ GeV 
				& 398
				& 500
				& 461
				& 1359        
				& 3285   
\\
$\pT{miss}>25$ GeV         
				& 3.2
                              	& 90
				& 357
				& 450         
      				& 2453        
\\
$|\mass{l^+l^-}-\mass{Z}|<10$ GeV 
				& 2.8
                                & 76
                                & 65
				& 144        
				& 2331 
\\
$\sumPtJet<100$ GeV        	
				& 2.5
                                & 72
                                & 44
				& 119        
				& 1987         
\\
\hline
\end{tabular}
\end{small}

\caption
{\label{t_selection_WZ} The number of events surviving after each of
the kinematic cuts is applied cumulatively for the $WZ$ analysis
for an integrated luminosity $\lumi=30~\mathrm{fb}^{-1}$.}
\end{table}

There are very few backgrounds which are able to mimic the leptonic $WZ$
signal. The most important backgrounds are (1) $ZZ$ production with leptonic
decays and one lepton escaping detection, and (2) $t\bar{t}$ production
with both of the $W$'s from the top quarks decaying leptonically and a
$b$-jet producing a third charged lepton. The contributions from each
background process are shown in Table~\ref{t_selection_WZ}.

The selection criteria for the $WZ$ analysis are: 
(1) three isolated electron or muons,
    $\pT{l^\pm}>25~\mathrm{GeV},~|\pseudorapitidy{l^\pm}|<2.5$
    two of which are like flavour, opposite sign leptons satisfying
    $|M(l^+,l^-)-\mass{Z}| < 10~\mathrm{GeV}$, 
(2) no other charged lepton with
    $\pT{l^\pm}>20~\mathrm{GeV},~|\pseudorapitidy{l^\pm}|<2.5$,
(3) missing transverse momentum $\pT{miss}>25~\mathrm{GeV}$,
(4) vector sum of jet transverse momenta (jet veto) 
   $\sumPtJet<100~\mathrm{GeV}$,
(5) a solution for the neutrino longitudinal momentum exists which 
   is consistent with it arising from a $W$.
About 2100 event candidates will be observed with an integrated luminosity 
of 30~fb$^{-1}$, 119 of which will be background.

\NLOa\ corrections degrade the \TGC\ sensitivity
because a large number of extra diagrams are included in the
calculation, the majority of which do not include the \TGC\ vertex.
The \NLOa\ corrections become largest when the jet activity is
significant. This means that a cut on the vector sum of the jet
transverse momenta $\sumPtJet$ will serve to moderate the influence of
these extra diagrams. When $\sumPtJet$ is small, the signal is
Born-like. When it is large, the diboson system will be recoiling
against a hard central jet, and the influence of the \TGC\ vertex will
be minimal.

The $\sumPtJet$ cut is optimised strictly on the basis of the
sensitivity to the \aTGC's. As the cut is increased, the signal purity
goes down, but at the same time the sensitivity increases until a
maximum is reached at about 100~GeV. This is because the signal itself
(in kinematic regions where the \aTGC's have little effect) is washing
out the sensitivity.

\section{Anomalous Coupling Confidence Limits}

\begin{figure}
\includegraphics[width=0.9\textwidth,height=0.45\textheight]{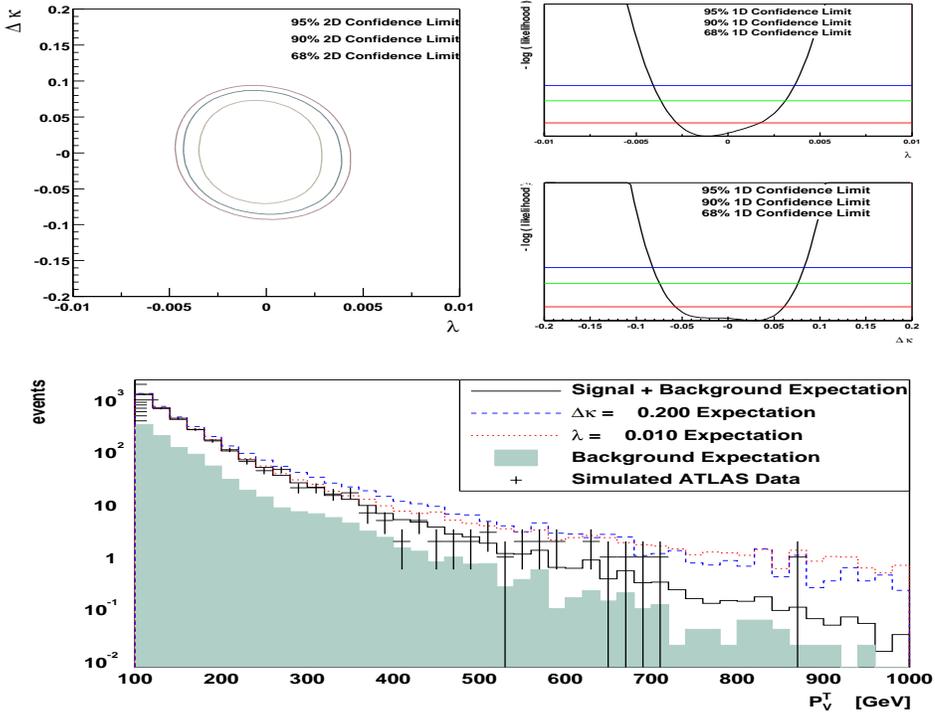}
\caption{
\label{f_ptV_WA}
The transverse momentum distribution of the photon in $W\gamma$
production is shown (bottom), together with the confidence intervals
(top) which may be extracted from this distribution.  }
\end{figure}

The expected statistical confidence intervals for \aTGC s are
evaluated by comparing histograms of `mock' \ATLAS\ data (simulated
with 30~fb$^{-1}$ and SM \TGC\ parameters) to reference histograms,
evaluated as a function of the \aTGC\ parameters, using a
binned maximum likelihood fit to one or two dimensional distributions.
As an example of the maximum likelihood fit, the transverse momentum
distribution of the photon in $W\gamma$ production is shown in
Fig.~\ref{f_ptV_WA}, after applying the kinematic cuts.  The points
with error bars represent ``mock'' data for one \ATLAS\ experiment.
The ``mock'' data histogram is constructed by sampling each bin
according to a Poisson distribution with the mean given by the
relevant bin content of the SM reference histogram.  The lines in
Fig.~\ref{f_ptV_WA} (bottom) are the reference distributions (i.e.\
theoretical expectation) for several choices of the \aTGC\ parameters.
The contribution of backgrounds to the reference distributions is
shown as a shaded histogram.  The one and two parameter negative log
likelihood curves are shown as a function of the $\lgamma$ and
$\dkgamma$ parameters with the 68, 90, and 95\% confidence limits
indicated. These confidence limits correspond to the single experiment
which has been simulated for this figure. When another \ATLAS\
experiment is simulated, the confidence limits will be different, on
account of statistical fluctuations.  In order to obtain the best
estimate of the limits that will be achieved, it is necessary to
average the confidence limits over many simulated \ATLAS\ experiments
(we use 5000 simulated experiments here).

When extracting the confidence intervals, the systematic uncertainties
are estimated by replacing the histograms which represent the `mock'
\ATLAS\ data with histograms which use a different model
assumption. The reference histogram assumptions are not changed.  The
change in the model assumptions causes a shift in the preferred value
for each \aTGC\ parameter.  This shift is independent of luminosity
and is taken as a pessimistic estimate of the systematic error, since
it is likely that it will be possible to extract corrections for many
of these systematic effects directly from the data. The following
systematic effects have been studied:
(1) Background rate systematics are evaluated by varying the
background process $k$-factor in the `mock' data histograms from 1.5
up to 2 and down to 1.
(2) Parton density function systematics are evaluated by replacing the
CTEQ4~\cite{Lai:1996mg} \pdf's which have been used for the all 
simulations with the CTEQ3~\cite{Lai:1994bb} series \pdf's in 
the `mock' data histograms.
(3) Systematics arising from neglected higher orders are evaluated by
varying the renormalisation and factorisation scales up and down by a
factor 2 for the $WZ$ signal simulation.
(4) Detector related systematics are evaluated by simply turning off
the detector smearing in the event generation software chain, which
represents the shift in the results that would arise if the \ATLAS\
detector were to be replaced by a fictional `perfect' detector.
Since the overall normalisation of the distributions does not enter into the maximum likelihood fit, uncertainties related to luminosity do not enter.
These systematic effects are uncorrelated and are added in quadrature
to obtain the total systematic error for the measurements.

The transverse momentum of the photon or $Z$-boson ($\pT{V}$) has been the
traditional means of extracting limits on the \aTGC's at hadron
colliders because it can be reconstructed without the assumptions necessary for
reconstructing the neutrino four-momentum and it projects out the
central, high diboson mass production regime where the \aTGC's are enhanced.

In addition to simple one dimensional distributions like the one shown
in Fig.~\ref{f_ptV_WA}, we have studied two-dimensional distributions
and have derived and applied a variation of the optimal observables
technique for hadron colliders (see \cite{Dobbs:2002WZ}).  In order to
produce reference histograms for two dimensional distributions in a
reasonable amount of computer time, the number of bins in each
dimension is reduced, which can reduce the sensitivity to these
distributions.

For the $\lz,\lgamma,~\mbox{and}~\dkgamma$ parameters, a maximum
likelihood fit to the one dimensional $\pT{V}$ distribution gives the
most stringent 95\% confidence intervals,
\begin{equation} \begin{array}{ccccc}
        \lzSystStatLimit \\
        \lgammaSystStatLimit \\
	\dkgammaSystStatLimit .
\end{array}
\end{equation}
The dominant systematic effects are theoretical, with the parton
density functions providing the biggest contribution to the $\lz$
limit, the modelling of QCD corrections being the biggest contribution
to the $\lgamma$ limit, and the background rate being the biggest
contribution to the $\dkgamma$ distribution.

The best 95\% confidence intervals for the $\dkz$ and $\dgOneZ$ parameters are obtained using the two dimensional $\pT{Z}$~vs.~$\pT{l_W}$
distributions,
\begin{equation} \begin{array}{ccccc}
        \dkzSystStatLimit    \\
	\dgOneZSystStatLimit 
\end{array} \end{equation}
The dominant systematic effect for these parameters comes from our theoretical
understanding of the proton structure (\pdf's). 

Since each \aTGC\ appears differently in the matrix elements, they
exhibit different sensitivity to each distribution. The $\lambda$-type
couplings, for example, appear proportional to energy squared and
$\sin \theta^*_V$, which makes them very sensitive to the $\pT{V}$
distribution. The $\kappa$-type couplings and $\dgOneZ$ are sensitive
to the vector-boson helicity, and so the transverse momentum of the
charged lepton from the $W^\pm$ (which acts like a projection
operator) is also a sensitive distribution.

For most of the \aTGC\ parameters, the confidence intervals are
dominated by statistics. This remains true for low luminosity
integrated luminosities in excess of 100~fb$^{-1}$. This is because
the sensitivity is dominated by the few events out in the high
transverse momentum tails, where the size of the event samples will
always be limited regardless of the total diboson event rate.  The
$\dgOneZ$ parameter sensitivity behaves very differently from the
other parameters, since this anomalous coupling parameter is more
sensitive to systematic effects, and a careful understanding and
evaluation of the systematic uncertainties will be particularly
important for measurements of this parameter at the \LHC. In addition
to \pdf's, detector related systematic effects will be of importance
for measurements of this parameter, particularly if two dimensional
distributions are employed.


An improvement in the statistical confidence intervals can be realized
for certain \TGC\ parameters (e.g.\ $\dkz$) by using the Optimal
Observable distributions (\OpOb) derived in Ref.~\cite{Dobbs:2002WZ}. However,
the calculation of the \OpOb's requires more in the way of
reconstruction and phenomenological input\footnote{To calculate the
\OpOb\ for a particular event, the centre-of-mass system needs to be
fully reconstructed such that all particle momenta are known, and
phenomenological parton density functions are included in the
calculation.}, and so these distributions are much more sensitive to
systematic effects. The systematic errors dominate the confidence
intervals for the \OpOb's, and no improvement in sensitivity is
realized. If the systematics can be controlled to a degree beyond what
has been assumed in this work, \OpOb's may provide a
viable means of measuring the \aTGC's.

\begin{figure}
  \includegraphics[width=0.5\textwidth,clip=]{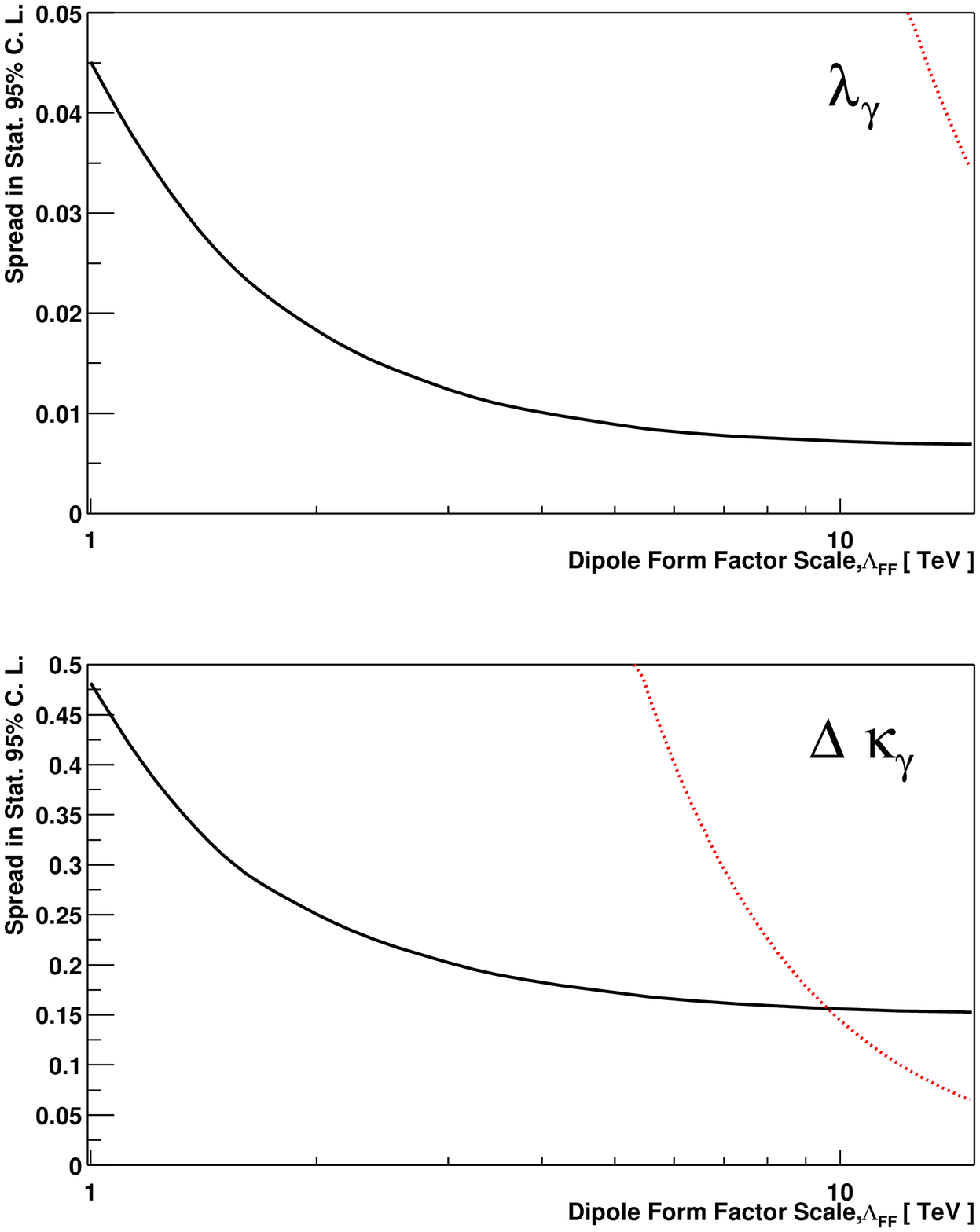}
  \includegraphics[width=0.5\textwidth,clip=]{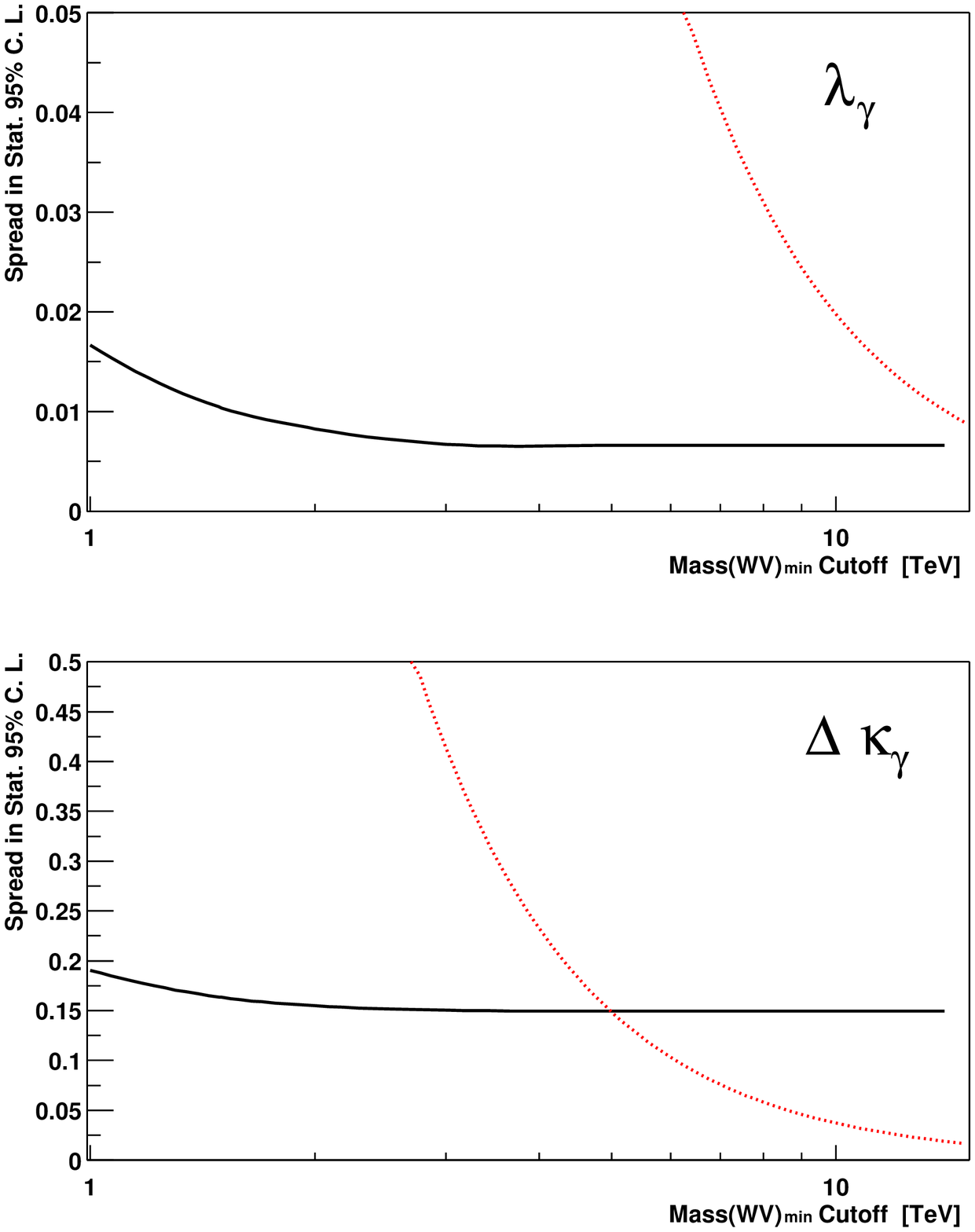}
\caption{\label{f_limits_vs_ff} The spread in statistical 95\%
  confidence intervals (solid lines) are shown as a function of the
  dipole form factor scale assumption $\LFF$ (left) and as a function
  the diboson mass cutoff (right) for the $\dkgamma$ parameter in
  $W\gamma$ production at the \LHC. The dotted lines indicate the
  approximate Born level unitarity limits.}
\end{figure}

For the results presented thus far, the \aTGC's have been assumed
constant, which would be in violation of unitarity at high energy
scales.  The most common approach for safe-guarding unitarity is to
multiply the anomalous couplings by a form factor
$(1+\frac{\mass{WV}^2}{\LFF^2})^{-n}$ (with $n=2$) which goes smoothly
to zero at high energy scales.  The form factor scale is usually
chosen to be so large for \TEVATRON\ analyses ($\LFF$=2~TeV) that the
effects of the form factor are not apparent at the scale at which the
experiment probes. As an example, the spread in the $\dkgamma$
confidence intervals are shown as a function of $\LFF$ in
Fig.~\ref{f_limits_vs_ff} (left). The limits improve with increasing $\LFF$ until
an asymptotic limit is reached at about 3-5~TeV, meaning the limits
presented above would not be degraded for $\LFF \ge 5$~TeV.

We advocate against an approach which uses a form factor scale which
is significantly smaller than the asymptotic $\LFF$ value (5~to 10~TeV
at the \LHC, and about 2~TeV at the \TEVATRON). The primary argument
in support of this philosophy is that the $\LFF$ defines the scale at
which the effective Lagrangian description (wherein the new physics
has been integrated out and described in terms of a small number of
low-dimensional operators) breaks down. Effectively, a scale has been
reached at which the effects of the new physics are directly
visible. There is no reason to expect the effects of this new physics
to turn off at that scale---rather it will appear directly, but will
not be parameterisable in terms of the effective \TGC\ Lagrangian. If
a form factor scale smaller than the asymptotic value is used, then it
will be absolutely essential to {\it neglect data} collected at the
scales where the assumed form factor operates. This is because in that
energy regime, the effective model Lagrangian is fully constrained to
the Standard Model (since the \aTGC's $\rightarrow$ 0 at $\LFF$), and
it makes no sense to include such data in a fit to extract the
\aTGC's.  However, the data which is collected at the largest scales
is potentially the most interesting, and one does not wish to be in a
position where it needs to be discarded.

We prefer to avoid the unnecessary dependence of experimental limits
on the two extra parameters ($n$ and $\LFF$) in the form factor choice
by reporting experimental \aTGC\ confidence limits as a function of
the diboson invariant mass being probed.  This is demonstrated in
Fig.~\ref{f_limits_vs_ff} (right), wherein the spread in the $\lgamma$
confidence intervals are shown as a function of a diboson invariant
mass cutoff (the minimum of the two reconstructed mass solutions is
used) which is applied to the data. For example, the limits at
Mass$(WV)_\txt{min}=2$~TeV use only the data for which the
reconstructed minimum mass solution is less than 2~TeV.  An asymptotic
limit is reached at about 3~TeV, meaning the \LHC\ is sensitive to
diboson masses up to about 3~TeV.  This treatment ensures unitarity,
without the need to introduce new parameters to parametrise the form
factor behaviour. The unitarity limit is superimposed on the plots as
a dotted line in Fig.~\ref{f_limits_vs_ff}.  The region above the
solid line is excluded by the experiment, while the region to the
right of the dotted line is excluded by unitarity. Reporting the
\aTGC\ limits as a function of the diboson mass makes the ultimate
mass reach of the experiment immediately evident, while allowing the
interpretation of results at any mass scale.  Further, if an anomalous
coupling `turns on' or `turns off' at some mass scale, that would be
reflected in the limits.

\begin{figure}
  \includegraphics[width=0.5\textwidth,height=0.35\textwidth]{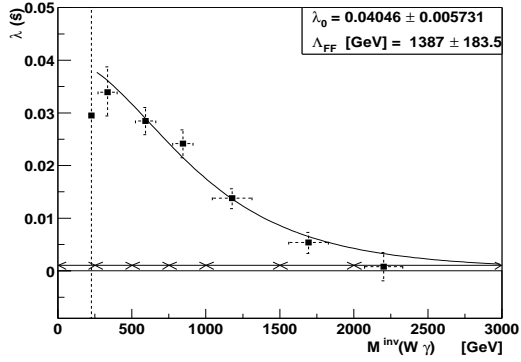}
 \caption{\label{f_limits_in_mass_bins}
          Measurement of the $\lgamma$ parameter as a function of energy
          is demonstrated using 30~fb$^{-1}$ integrated luminosity for
          $W\gamma$ production.  The `mock' \ATLAS\ data has been
          generated with ${\lgamma}_0$=0.04 using a dipole form factor of
          scale $\LFF=1500$~GeV. The solid line is a fit to $\LFF$ and
          the bare coupling ${\lz}_0$ assuming the dipole form
          factor. The arrows along the $x$-axis indicate the diboson
          mass bin widths.  }
\end{figure}

In the scenario where \aTGC\ measurements at \LHC\ are inconsistent
with the Standard Model, it would be preferable to measure the energy
dependence of the \aTGC\ parameters directly, rather than assuming
some energy dependence in the model.
A large data sample of diboson events will be necessary to perform
such a measurement, because the data needs to be separated out into
bins of diboson mass.  For Fig.~\ref{f_limits_in_mass_bins}, `mock'
\ATLAS\ data has been generated with bare coupling ${\lgamma}_0=0.04$ and
a dipole ($n$=2) form factor with $\LFF=1500$~GeV. This `mock data' is
then compared to reference histograms of the bare coupling
${\lgamma}_0$ (i.e.\ the reference histograms do not use a form
factor) for each of the diboson mass bins. The events have been
separated out into diboson mass bins ranging from 250~GeV to 3000~GeV
with variable width, to ensure adequate statistics in each bin. 
The behaviour of the couplings as a function of energy is clearly
visible. A fit to the dipole form factor function is also indicated
with a solid line. The parameters which were used to generate the
`mock' data are reproduced within the precision of the fit.

\section{Radiation Zero in $W\gamma$ Production}

The radiation zero refers to a particular center-of-mass frame
emission angle of the photon with respect to the anti-quark ($\mycos$) in $W\gamma$ production which is
forbidden by subtle gauge cancellations (an approximate radiation zero
exists for $WZ$ production). The radiation zero has yet to be observed
experimentally. 

\begin{figure}
  \includegraphics[width=.5\textwidth,height=.3\textwidth]{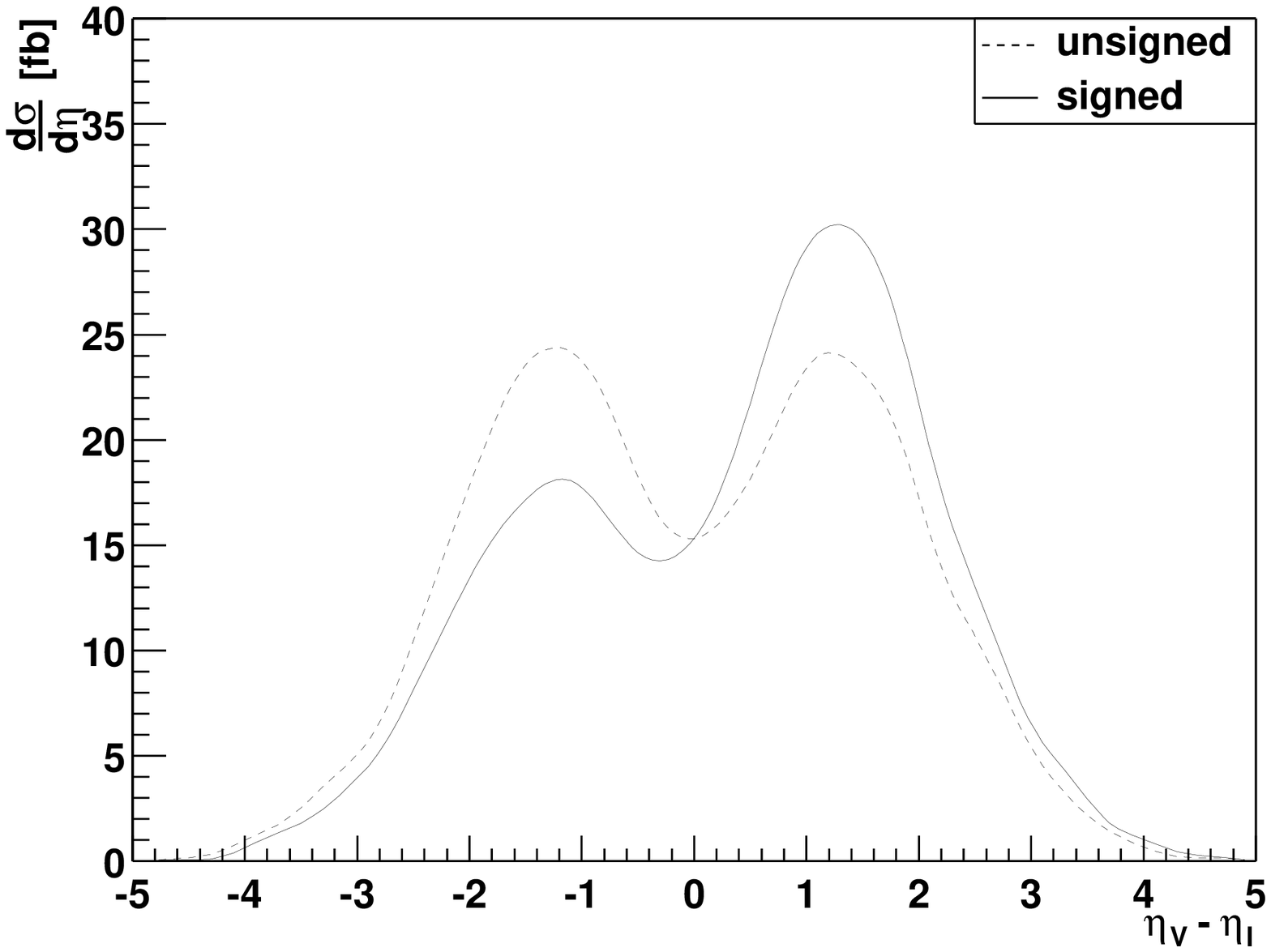}
  \includegraphics[width=.5\textwidth,clip=]{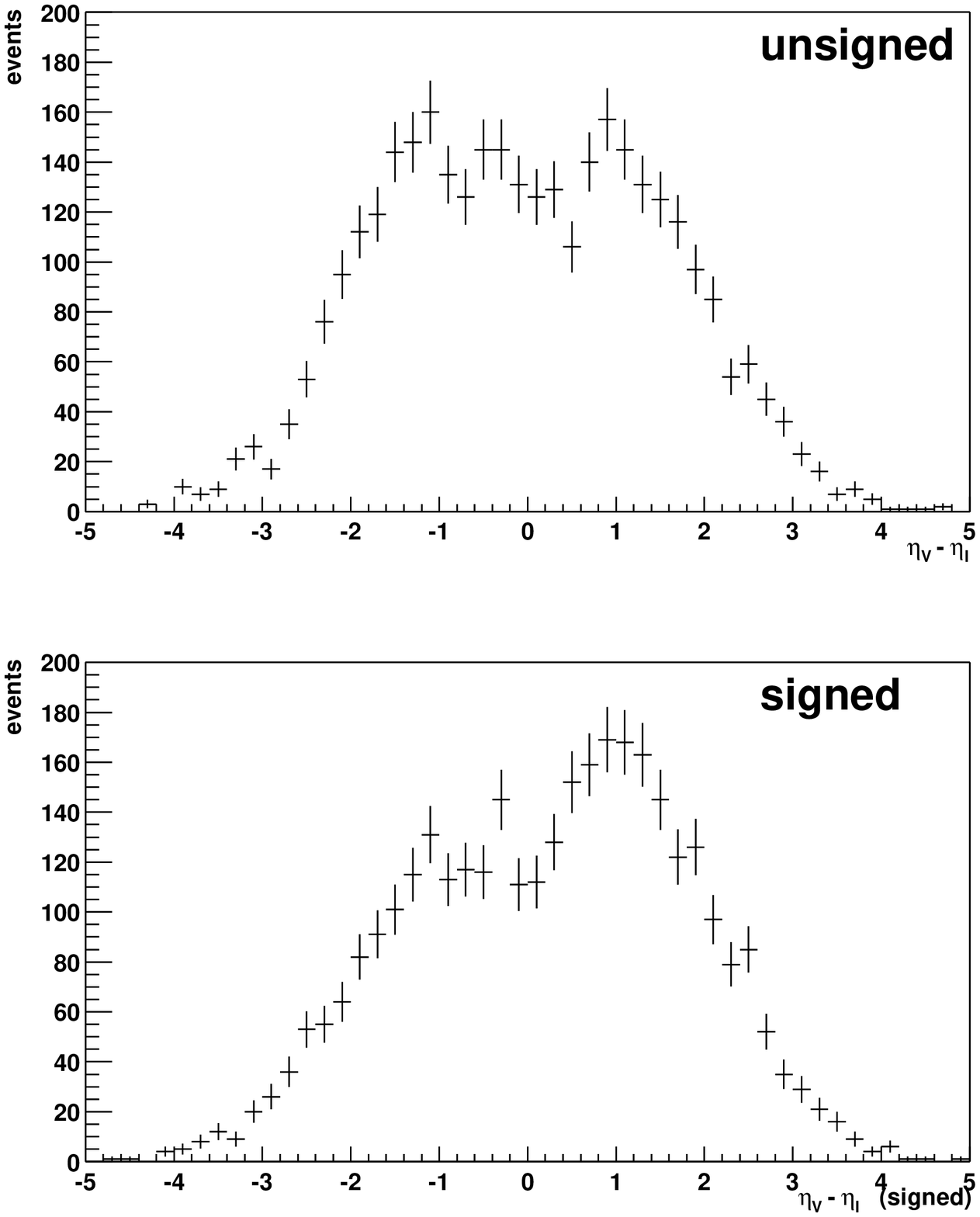}
  \caption{\label{f_eta_separation_signed_WA_THY} The rapidity
  separation of the $\gamma$ from the $l^\pm_W$ is shown for $W\gamma$
  production at the \LHC.  For the solid lines in the theoretical
  distribution on the left, the rapidity separation has been `signed'
  according to the overall boost of the event. The distribution for
  30~fb$^{-1}$ of \LHC\ data is shown right.  }
\end{figure}

For hadronic collisions, the pseudorapidity difference between the
photon and charged lepton
$\pseudorapitidy{\gamma}-\pseudorapitidy{l^\pm_W}$ is normally used to
demonstrate the effects of the radiation zero. For $p\bar{p}$
collisions, this distribution shows the characteristic antisymmetric
shape which many people have claimed gives $p\bar{p}$ experiments an
advantage over $pp$ experiments.  For symmetric proton-proton
collisions, it is not possible to ascertain from which beam the quark
or antiquark arises, and this washes out the radiation zero, such that
it shows up only as a small dip at $\eta=0$.  The idea of ``signing''
the quark direction according to the overall boost of each event was
introduced in Ref.~\cite{Fisher:1994pw,Baur:1997wa} in the context of
measuring the electroweak mixing angle with dilepton events.  We
apply that idea to diboson production in
Fig.~\ref{f_eta_separation_signed_WA_THY} (left), and find that the
characteristic asymmetric radiation zero shape is recovered.

The effect will be clearly observable with
30~fb$^{-1}$ of data from the \LHC\ as shown in Fig.~\ref{f_eta_separation_signed_WA_THY} (right), wherein the signed $\eta$ distribution is plotted for SM $W\gamma$ production.

\section{Conclusions}

The prospects for measuring the $WW\gamma$ and $WWZ$ \TGC\ vertex at
the \LHC\ have been assessed in the context of the \ATLAS\
experiment. The leptonic decay channels of $W\gamma$ and $WZ$ diboson
production provide clean signatures with signal to background ratios
of 1.6 and 17 respectively. For the majority of the \aTGC\ parameters,
the confidence intervals will be dominated by statistics for
integrated luminosities up to and beyond 100~fb$^{-1}$. For the $\dgOneZ$
parameter, our ability to model the proton structure with the parton
density functions will be a challenging systematic.

Dipole form factors have been the conventional means of guaranteeing
unitarity in the \TGC\ Lagrangian. The parametrisation of the form
factors is arbitrary, and introduces unnecessary dependence on the
parametrisation choice into the experimental results. We have argued
that it is preferable to report the limits as a function of a diboson
invariant mass cutoff which is applied to the data.  The \LHC\ will
directly probe diboson invariant masses up to about 3~TeV.  In this
mass regime, the limits reported in this paper are unitarity safe and
are presented without any cutoff of form factor. In the scenario where
non-standard \aTGC\ parameters are observed, the \LHC\ event rate will
be sufficiently large to bin the data according to the diboson
invariant mass, and measure the energy dependence of the couplings
directly.

The expected \LHC\ confidence limits for 30~fb$^{-1}$ (including
systematic effects) are $\lzCombinedLimit,~ \dkzCombinedLimit,~
\dgOneZCombinedLimit,~
\lgammaCombinedLimit,~\mbox{and}~\dkgammaCombinedLimit$.

%
%
%
%


\begin{theacknowledgments}

I thank the HCP organisers for providing a fun and stimulating
environment. I thank the many ATLAS and CMS collaborators who
contributed to the analyses presented in my HCP talk. Michel Lefebvre
was a valued collaborator for the ATLAS $WZ$ and $W\gamma$ \TGC\
studies presented in these proceedings.  This work was supported in
part by the Director, Office of Science, Office of Basic Energy
Sciences, of the U.S. Department of Energy under Contract
No. DE-AC03-76SF00098.

\end{theacknowledgments}


\bibliographystyle{aipproc}   

\bibliography{2004-HCP-Mdobbs}

\begin{thebibliography}{28}
\expandafter\ifx\csname natexlab\endcsname\relax\def\natexlab#1{#1}\fi
\providecommand{\enquote}[1]{``#1''}
\expandafter\ifx\csname url\endcsname\relax
  \def\url#1{\texttt{#1}}\fi
\expandafter\ifx\csname urlprefix\endcsname\relax\def\urlprefix{URL }\fi

\bibitem[Atl(1999)]{Atlas:TDR}
{ATLAS Detector and Physics Performance Technical Design Report}, Tech. rep.,
  CERN/LHCC/99-15 (1999).

\bibitem[Marques et~al.(2005)]{marques:2004Wmass}
Marques, C., Maio, A., Pallin, D., and Dobbs, M., \enquote{{W mass measurement
  with Atlas},} in \emph{Physics at LHC 13-17 July 2004, Vienna, Austria},
  2005.

\bibitem[Sliwa et~al.(2000)]{Sliwa:2000}
Sliwa, K., Riley, S., and Baur, U., {Effects of possible extensions to rapidity
  coverage of the ATLAS detector on the determination of
  $\sin^2\theta_{eff}(M^2_Z)$}, Tech. rep., CERN-ATL-PHYS-2000-018 (2000).

\bibitem[Bell(2003)]{Bell:2003phd}
Bell, P., \emph{{Anomalous quartic couplings at OPAL and the system test of the
  ATLAS barrel SCT}}, Ph.D. thesis, Birmingham University, Birmingham, U.K.
  (2003).

\bibitem[Dobbs et~al.(2004)]{Dobbs:2004qw}
Dobbs, M., Frixione, S., Laenen, E., and Tollefson, K., editors, \emph{Les
  Houches guidebook to Monte Carlo generators for hadron collider physics},
  2004, hep-ph/0403045.

\bibitem[Dobbs and Lefebvre(2002{\natexlab{a}})]{Dobbs:2002WZ}
Dobbs, M., and Lefebvre, M., {Prospects for Probing the Three Gauge-boson
  Couplings in W + Z Production at the LHC}, Tech. rep., CERN-ATL-PHYS-2002-023
  (2002{\natexlab{a}}).

\bibitem[Dobbs and Lefebvre(2002{\natexlab{b}})]{Dobbs:2002WA}
Dobbs, M., and Lefebvre, M., {Prospects for Probing the Three Gauge-boson
  Couplings in W + Photon Production at the LHC}, Tech. rep.,
  CERN-ATL-PHYS-2002-019 (2002{\natexlab{b}}).

\bibitem[Dobbs(2002)]{Dobbs:2002phd}
Dobbs, M., \emph{{Probing the Three Gauge-boson Couplings in 14-TeV Proton
  Proton Collisions}}, Ph.D. thesis, University of Victoria, Victoria, Canada
  (2002), uMI-NQ-68128.

\bibitem[Mackay(1998)]{mackay:1998phd}
Mackay, C., \emph{{The Electromagnetic Calorimeter For CMS and a Study of the W
  W gamma Vertex}}, Ph.D. thesis, Brunel University (1998).

\bibitem[M\"uller et~al.(2000)]{Muller:2000tgc}
M\"uller, T., Neuberger, D., and Th\"ummel, W., {Sensitivities for anomalous
  WWgamma and ZZgamma couplings at CMS}, Tech. rep., CERN-CMS-NOTE-2000-017
  (2000).

\bibitem[Mackay(2001)]{Mackay:2001tgc1}
Mackay, C., {Observing anomalous di-boson couplings in the WWgamma vertex at
  CMS}, Tech. rep., CERN-CMS-NOTE-2001-052 (2001).

\bibitem[Mackay and Hobson(2001)]{Mackay:2001tgc2}
Mackay, C., and Hobson, P., {Sensitivity of CMS to CP conserving anomalous
  di-boson couplings in Wgamma events}, Tech. rep., CERN-CMS-NOTE-2001-056
  (2001).

\bibitem[Barate et~al.(1999)]{Barate:1999gu}
Barate, R., et~al., \emph{Phys. Lett.}, \textbf{B453}, 107--120 (1999).

\bibitem[Abreu et~al.(1999)]{Abreu:1999ra}
Abreu, P., et~al., \emph{Phys. Lett.}, \textbf{B456}, 310--321 (1999).

\bibitem[Acciarri et~al.(1998)]{Acciarri:1998aq}
Acciarri, M., et~al., \emph{Phys. Lett.}, \textbf{B436}, 437--452 (1998).

\bibitem[Abbiendi et~al.(1999)]{Abbiendi:1998bg}
Abbiendi, G., et~al., \emph{Eur. Phys. J.}, \textbf{C8}, 191--215 (1999).

\bibitem[Mackay and Hobson(2002)]{Mackay:2002tgc}
Mackay, C., and Hobson, P., {Observing Anomalous Di-boson couplings in Z gamma
  gamma and Z Z gamma vertices at CMS}, Tech. rep., CERN-CMS-NOTE-2002-028
  (2002).

\bibitem[Hassani(2002{\natexlab{a}})]{Hassani:2002ZZ}
Hassani, S., {Prospects for Measuring Neutral Gauge Boson Couplings in $ZZ$
  Production with the ATLAS Detector}, Tech. rep., CERN-ATL-PHYS-2002-023
  (2002{\natexlab{a}}).

\bibitem[Hassani(2002{\natexlab{b}})]{Hassani:2002ZA}
Hassani, S., {Prospects for Measuring Neutral Gauge Boson Couplings in
  $Z\gamma$ Production with the ATLAS Detector}, Tech. rep.,
  CERN-ATL-PHYS-2002-012 (2002{\natexlab{b}}).

\bibitem[Richter-Was et~al.(1998)]{RichterWas:1998atlfast}
Richter-Was, E., Froidevaux, D., and Poggioli, L., {ATLFAST 2.0 a fast
  simulation package for ATLAS}, Tech. rep., CERN-ATL-PHYS-98-131 (1998).

\bibitem[Baur et~al.(1993)]{Baur:1993ir}
Baur, U., Han, T., and Ohnemus, J., \emph{Phys. Rev.}, \textbf{D48}, 5140--5161
  (1993).

\bibitem[Baur et~al.(1995)]{Baur:1994aj}
Baur, U., Han, T., and Ohnemus, J., \emph{Phys. Rev.}, \textbf{D51}, 3381--3407
  (1995).

\bibitem[Sjostrand et~al.(2001)]{Sjostrand:2001wi}
Sjostrand, T., et~al., \emph{Comput. Phys. Commun.}, \textbf{135}, 238--259
  (2001).

\bibitem[Kelly(1996)]{kelly96}
Kelly, M., \emph{{Test of the standard model of electroweak interactions by
  measuring the anomalous $WW\gamma$ coupling at $\sqrt{s}=1.8$~TeV}}, Ph.D.
  thesis, Notre Dame, South Bend Indiana (1996).

\bibitem[Lai et~al.(1997)]{Lai:1996mg}
Lai, H.~L., et~al., \emph{Phys. Rev.}, \textbf{D55}, 1280--1296 (1997).

\bibitem[Lai et~al.(1995)]{Lai:1994bb}
Lai, H.~L., et~al., \emph{Phys. Rev.}, \textbf{D51}, 4763--4782 (1995).

\bibitem[Fisher et~al.(1995)]{Fisher:1994pw}
Fisher, P., Becker, U., and Kirkby, J., \emph{Phys. Lett.}, \textbf{B356},
  404--408 (1995).

\bibitem[Baur et~al.(1998)]{Baur:1997wa}
Baur, U., Keller, S., and Sakumoto, W.~K., \emph{Phys. Rev.}, \textbf{D57},
  199--215 (1998).

\end{thebibliography}

\IfFileExists{\jobname.bbl}{}
 {\typeout{}
  \typeout{******************************************}
  \typeout{** Please run "bibtex \jobname" to optain}
  \typeout{** the bibliography and then re-run LaTeX}
  \typeout{** twice to fix the references!}
  \typeout{******************************************}
  \typeout{}
 }

\end{document}